\documentclass[10pt, conference, compsocconf]{IEEEtran}
\usepackage{cite}
\usepackage{amsmath,amssymb,amsfonts}
\usepackage{algorithmic}
\usepackage[pdftex]{graphicx}
\usepackage{textcomp}
\usepackage{xcolor}
\def\BibTeX{{\rm B\kern-.05em{\sc i\kern-.025em b}\kern-.08em
    T\kern-.1667em\lower.7ex\hbox{E}\kern-.125emX}}

\usepackage{tikz,lipsum}
\usepackage[most]{tcolorbox}
\AtBeginDocument{%
  \providecommand\BibTeX{{%
    \normalfont B\kern-0.5em{\scshape i\kern-0.25em b}\kern-0.8em\TeX}}}

\usepackage[utf8]{inputenc}
\usepackage{amsmath}
\usepackage{amssymb}
\usepackage{hyperref}
\usepackage{balance}
\usepackage{url}
\usepackage{multirow}
\usepackage{caption}
\usepackage{wrapfig}
\usepackage{color,soul}
\usepackage{fancyvrb}
\usepackage{eepic}
\usepackage{enumitem}
\usepackage[T1]{fontenc}
\usepackage[scaled=0.85]{beramono}
\usepackage{listings}
\usepackage{balance}
\usepackage{booktabs}
\usepackage{array}
\newcolumntype{P}[1]{>{\centering\arraybackslash}p{#1}}

\def\shorten{\looseness=-1} 
\usepackage{makecell}

\def\ncp{\vspace*{-0.5ex}}
\makeatletter
\newif\if@restonecol
\makeatother

\usepackage[ruled,vlined,linesnumbered]{algorithm2e}

\newcommand{\sysName}{KGNet} 

\usepackage[hyphenbreaks]{breakurl}

\newcommand{\RDFTYPE}[3]{\ensuremath{{\langle}\texttt{#1},} \texttt{#2}, \ensuremath{\texttt{#3}{\rangle}}}

\newcommand{\myNum}[1]{(\emph{#1})}

\SetKwComment{Comment}{{// }}{}

\def\ncp{\vspace*{-0.5ex}}
\begin{document}

\title {Towards a GML-Enabled Knowledge Graph Platform\shorten}

\author{\IEEEauthorblockN{ Hussein Abdallah}
\IEEEauthorblockA{\textit{Concordia University, }\\
hussein.abdallah@concordia.ca}
\and
\IEEEauthorblockN{Essam Mansour}
\IEEEauthorblockA{\textit{Concordia University, }\\
essam.mansour@concordia.ca}
}


\maketitle
\begin{abstract}

This vision paper proposes {\sysName}, an on-demand graph machine learning (GML) as a service on top of RDF engines to support GML-enabled SPARQL queries. {\sysName} automates the training of GML models on a KG by identifying a task-specific subgraph. This helps reduce the task-irrelevant KG structure and properties for better scalability and accuracy. While training a GML model on $KG$, {\sysName} collects metadata of trained models in the form of an RDF graph called KGMeta, which is interlinked with the relevant subgraphs in $KG$. Finally, all trained models are accessible via a SPARQL-like query. We call it a GML-enabled query and refer to it as SPARQL$^\mathbf{ML}$. {\sysName} supports SPARQL$^\mathbf{ML}$ on top of existing RDF engines as an interface for querying and inferencing over KGs using GML models. The development of {\sysName} poses research opportunities in several areas, including meta-sampling for identifying task-specific subgraphs, GML pipeline automation with computational constraints, such as limited time and memory budget,  and SPARQL$^\mathbf{ML}$ query optimization. {\sysName} supports different GML tasks, such as node classification, link prediction, and semantic entity matching. 
We evaluated  {\sysName} using two real KGs of different application domains. Compared to training on the entire KG, {\sysName} significantly reduced training time and memory usage while maintaining comparable or improved accuracy. The {\sysName} source-code\footnote{\url{https://github.com/CoDS-GCS/KGNET}} is available for further study. 
\end{abstract}

\section{Introduction}
\label{sec:intro} 

Knowledge graphs (KGs) are constructed based on semantics captured from heterogeneous datasets using various Artificial Intelligence (AI) techniques, such as representation learning and classification models~\cite{Knowledge_Graphs_Hogan}. 
Graph machine learning (GML) techniques, such as graph representation learning and graph neural networks (GNNs), are powerful tools widely used to solve real-world problems by defining them as prediction tasks on KGs. For instance, node classification tasks for problems, such as recommendations~\cite{gnnRS} and entity alignment~\cite{gnnCrossLingual}, can be solved using GML techniques. Similarly, drug discovery~\cite{kgnn} and fraud detection~\cite{fraudDet, CARE-GNN} problems are tackled as link prediction tasks using GML techniques.\shorten 

Data scientists often work with KGs, which are typically stored in RDF engines. They are responsible for developing GML pipelines using frameworks, such as PyG~\cite{pytorch-geometric} and DGL~\cite{DistDGL}, to train models on these KGs. However, there is often a gap between the GML frameworks and RDF engines. This necessitates an initial step of transforming the entire KG from RDF triple format into adjacency matrices in a traditional GML pipeline. Afterward, the data scientist needs to select a suitable GML method from a wide range of KG embedding (KGE) or GNN methods~\cite{GNN_Survey_2021, KG-Embedding-Survey} to train the model. For the average user, this responsibility is time-consuming. Furthermore, the trained models are isolated from the RDF engine, where the KG is stored. Therefore, automating the training of GML models on KGs and providing accessibility to the trained models via a SPARQL-like query is essential. We refer to this query as a SPARQL$^\mathbf{ML}$ query.\shorten

\begin{figure}[t]
\centering
  \includegraphics[width=\columnwidth,draft=false]{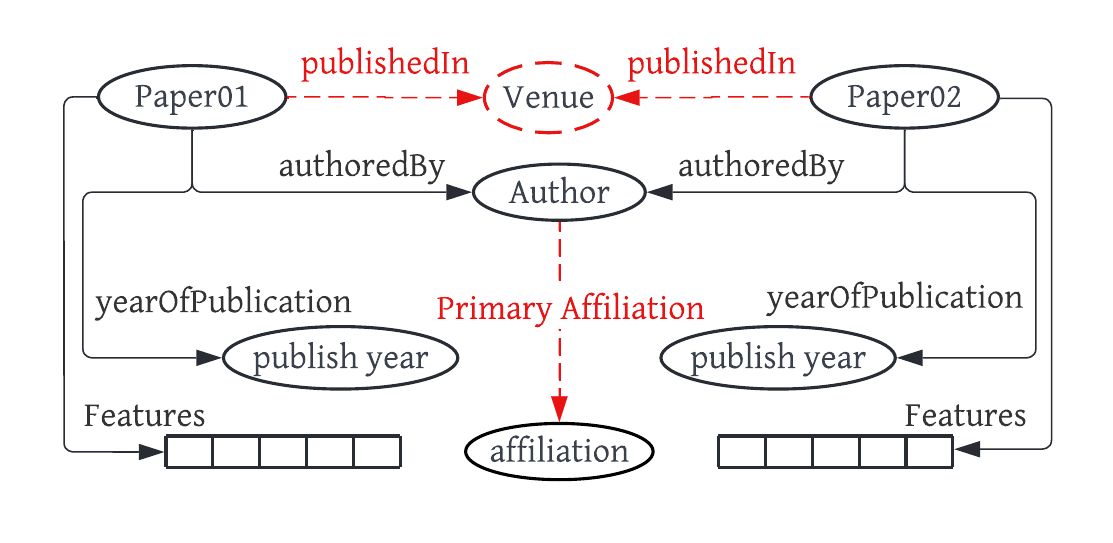}  
  \vspace*{-4ex}
  \caption{A KG with nodes/edges in red, which could be predicted by classification and link prediction models on the fly.\shorten}
  \label{fig:Paper_Venue_KG}
  \vspace*{-2ex}
\end{figure}

\begin{figure}[t]
\lstdefinestyle{myStyle}{basicstyle=\small\ttfamily, language=SPARQL }
\begin{lstlisting}[style=myStyle,captionpos=b,showspaces=false,frame=singlem,numbers=left,xleftmargin=0.5cm,morekeywords={kgnet, getNodeClass,dblp},escapechar=\%]
prefix dblp: <https://www.dblp.org/>
prefix kgnet: <https://www.kgnet.com/>
select ?title ?venue
where { 
?paper a dblp:Publication.
?paper dblp:title ?title.
?paper %\textbf{?NodeClassifier}% ?venue.
?NodeClassifier a %\textcolor{purple}{ \textbf{kgnet:NodeClassifier}}%.
?NodeClassifier kgnet:TargetNode dblp:Publication.
?NodeClassifier kgnet:NodeLabel dblp:venue.}
\end{lstlisting}
\ncp\ncp
\caption{SPARQL$^\mathbf{ML}_{pv}$: a SPARQL$^\mathbf{ML}$ query uses a node classification model to predict a paper's venue by querying and inferencing over the KG shown in Figure~\ref{fig:Paper_Venue_KG}.\shorten}
\label{fig:paperVenueSPARQL}
\ncp\ncp
\end{figure}

The KG shown in Figure~\ref{fig:Paper_Venue_KG} contains information about published papers in DBLP~\cite{KG_DBLP}. However, the traditional SPARQL query language cannot be used to apply GML models on top of a KG, such as predicting a node's class or a missing affiliation link for an author. For instance, the venue node in Figure~\ref{fig:Paper_Venue_KG} is a virtual node that could be predicted using a node classification (NC) model. It would be fascinating to query this KG using a GML model for NC through a SPARQL-like query to obtain the paper-venue node, as shown in the SPARQL$^\mathbf{ML}_{pv}$ in Figure~\ref{fig:paperVenueSPARQL}. This query uses a model of type \emph{kgnet:NodeClassifier} to predict a venue for each paper.
The SPARQL$^\mathbf{ML}$ triple patterns in lines 8-10 will retrieve all models of type \emph{kgnet:NodeClassifier} that predict a class of type \emph{dblp:venue}. In the triple pattern \RDFTYPE{?paper}{?NodeClassifier}{?venue}, we refer to \emph{?NodeClassifier} as a user-defined predicate.\shorten

Enabling queries like SPARQL$^\mathbf{ML}_{pv}$, shown in Figure~\ref{fig:paperVenueSPARQL}, presents several challenges. These include: \myNum{i} automatically training GML models for various tasks, \myNum{ii} optimizing SPARQL$^\mathbf{ML}$ for GML model selection based on accuracy and inference time, and \myNum{iii} efficiently interacting with the selected model during query execution. Additionally, seamless integration of GML models into RDF engines is necessary. As a result, users should be able to express their SPARQL$^\mathbf{ML}$ queries easily by following the SPARQL logic of pattern matching, avoiding the explicit use of user-defined functions (UDFs).\shorten

There is a growing adoption of integrating GML with existing graph databases, such as Neo4j~\cite{Neo4j_semantic_similarity} or Stardog~\cite{Stardog-ML}. However, while these databases offer some machine learning primitive methods, such as PageRank and shortest-path using the \emph{Cypher} language, they do not address the challenges of integrating GML models with RDF engines.
For example, Neo4j Graph Data Science~\cite{Neo4j_DS} supports limited graph embedding methods in a beta version, such as FastRP~\cite{FastRP}, Node2Vec~\cite{Node2Vec}, and Graph-SAGE~\cite{Graph-SAGE}. However, a user must train the models separately as an initial step.
To address these challenges, there is a need to bring GML to data stored in RDF engines instead of getting data to machine learning pipelines. This would encourage the development of KG data science libraries powered by the expressiveness of SPARQL, enabling better analysis and insight discovery based on KG structure and semantics. These libraries would empower data scientists with a full breadth of KG machine learning services on top of KGs stored in RDF engines.\shorten

\begin{figure}[t]
\ncp\ncp
  \centering
  \includegraphics[width=\columnwidth,draft=false]{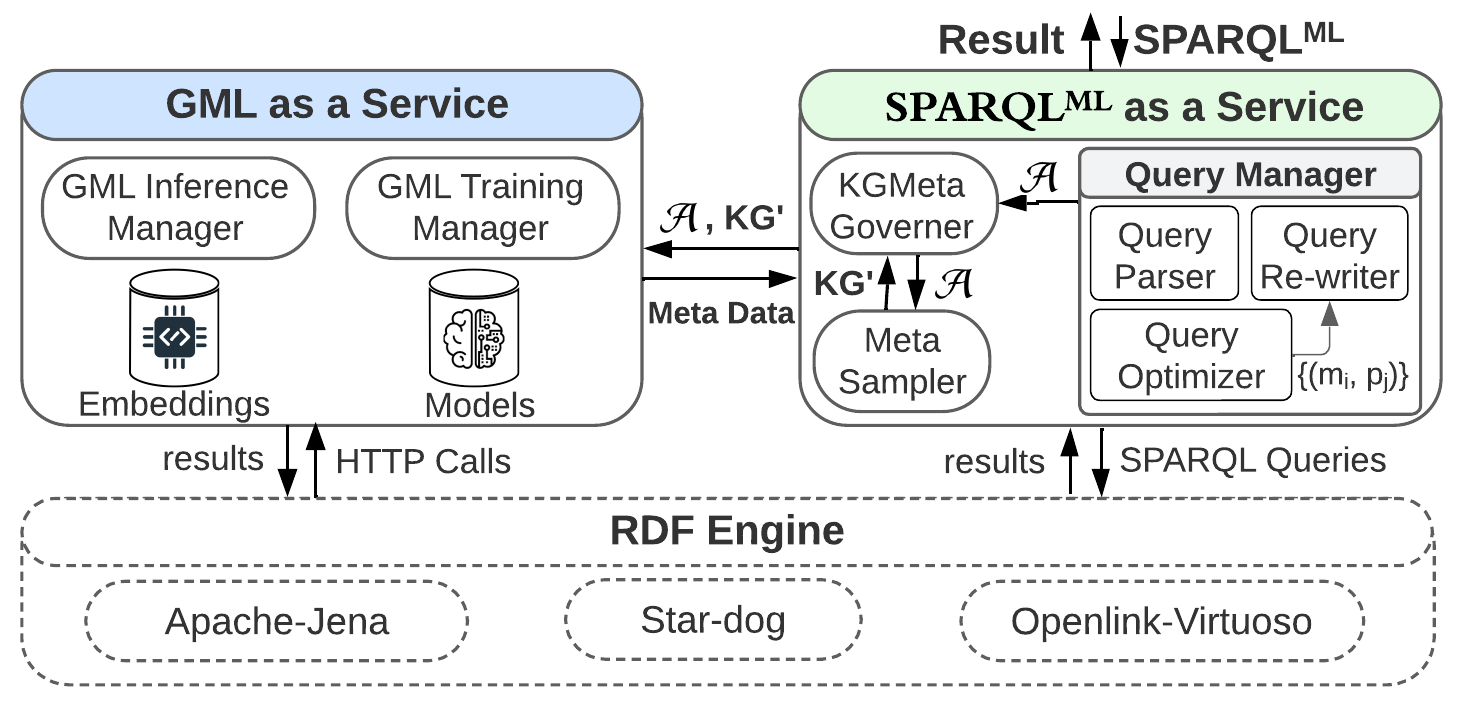}
  \caption{The {\sysName} architecture, which provides an interface language (SPARQL$^\mathbf{ML}$) and enables AI applications and data scientists to automatically train GML models  on top of KGs for querying and inferencing KGs based on the trained models.}  
  \ncp\ncp\ncp
  \label{fig:sys}
  \end{figure}

This vision paper proposes {\sysName}, an on-demand GML-as-a-service on top of RDF engines to support SPARQL$^\mathbf{ML}$ queries, as illustrated in Figure~\ref{fig:sys}. {\sysName} extends existing RDF engines with two main components GML-as-a-service (GMLaaS) and SPARQL$^\mathbf{ML}$ as a Service.
{\sysName} automatically trains a GML model on a KG for tasks, such as node classification or like prediction, and maintains metadata of the trained model as an RDF graph called \emph{KGMeta}. To reduce training time and memory usage while improving accuracy on a specific task $\mathcal{A}$, {\sysName} performs meta-sampling to identify a task-specific subgraph $KG'$ of the larger KG that preserves essential characteristics relevant to $\mathcal{A}$. This enables {\sysName} to scale on large KGs.  
GMLaaS is in charge of: \myNum{i} selecting the near-optimal GML method for training $\mathcal{A}$ using $KG'$ based on a given time or memory budget, and \myNum{ii} communicating with RDF engines via HTTP calls requesting inferencing of a specific trained model, \myNum{iii} storing the trained models and embeddings related to KGs. The SPARQL$^\mathbf{ML}$ service transparently: \myNum{i} maintains and interlinks the KGMeta with associated KGs, \myNum{ii} optimizes the GML model selection for a user-defined predicate, and \myNum{iii} finally rewrites the SPARQL$^\mathbf{ML}$ query as a SPARQL query.\shorten

In summary, the contributions of this paper are:
\begin{itemize}
\item a fully-fledged GML-enabled KG platform\footnote{\url{https://github.com/CoDS-GCS/KGNET}}  on top of existing RDF engines.\shorten
 \item GML-as-a-service to provide automatic training of GML models based on a given memory or time budget. This automatic training utilizes task-specific subgraphs extracted using our meta-sampling approach.
\item SPARQL$^\mathbf{ML}$ as a Service to perform meta-sampling, maintain training meta-data in KGMeta, and optimize the GML model selection, i.e., opt for the near-optimal model based on  constraints on accuracy and inference time. 
\item A comprehensive evaluation with different GML methods using three GML tasks on real KGs.  Our experiments show that {\sysName} achieved comparable or improved accuracy compared to training on the entire KG, while significantly reducing training time and memory usage.
\end{itemize}

The remainder of this paper is organized as follows. 
Section ~\ref{sec:GML-pipeline-for-KG} provides a background about existing graph machine learning pipelines. 
Section~\ref{sec:ChallengesOfAutomatingCognitiveQuery} outlines the main research challenges of developing  a GML-enabled KG engine. 
Section~\ref{sec:arch} presents the {\sysName} platform. 
Section~\ref{sec:eval} discusses the results of evaluating our automated pipeline for training GML models.
Sections~\ref{sec:related_work} and ~\ref{sec:conc} are related work and conclusion.


\section{Background: ML pipelines for KGs}
 \label{sec:GML-pipeline-for-KG}
 
ML pipelines developed to train models on a KG can be grouped into three main categories: \myNum{i} traditional ML on KG data in tabular format, \myNum{ii} traditional ML on KG embeddings, and \myNum{iii} graph neural networks (GNNs) trained directly on the KG.
In the traditional ML approach using KG data in tabular format, data from the KG is transformed into in-memory data frames, and classical ML classifiers are trained using feature engineering techniques and libraries, such as Scikit-Learn or SparkMLib. In contrast, traditional ML on KG embeddings avoids the feature engineering process and generates embeddings for nodes and edges. Apple Saga\cite{Saga_Apple} is an example of this approach, which uses graph ML libraries like DGL-KE \cite{DistDGL} to generate KG embeddings. Data scientists have the flexibility to choose the ML method for training.

\begin{figure}[t]
\ncp\ncp
  \centering  \includegraphics[width=0.8\columnwidth,draft=false]{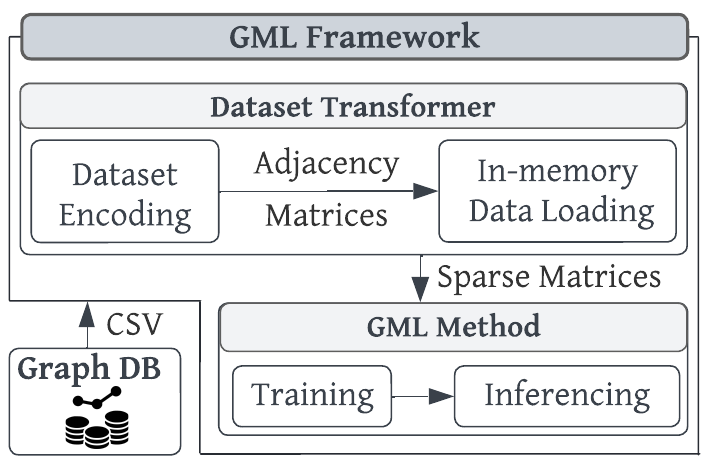}
  \ncp\ncp
  \caption{A traditional GML pipeline~\cite{OGB} using a GML framework. The pipeline starts with extracting the graph data, followed by data transformation into sparse matrices to train models for a GML task. Finally, the inference step is ready to predict results in isolation from the graph databases.}
  \ncp\ncp
  \label{fig:GNN_Pipeline} 
\end{figure}

GNNs have gained significant popularity in recent years. Hence, data scientists frequently utilize them to perform GML tasks. The Open Graph Benchmark (OGB)~\cite{OGB} standardized the GNN training pipeline, emphasizing the best practices for tackling GML tasks and building a GNN training pipeline. Figure~\ref{fig:GNN_Pipeline} summarizes this pipeline, which involves encoding KG nodes and edges, generating adjacency matrices, loading them into memory, and training GNNs using specific methods.

Various GML frameworks, such as DGL \cite{DistDGL} and PyG \cite{pytorch-geometric}, offer multiple implementations of GNN methods. These frameworks support data transformation by loading graphs into memory as graph data structures and applying transformations. However, existing GML frameworks require significant memory and processing time for large KGs and a deep understanding of various GNN methods. In comparison, the OGB pipeline is simple, but it is a semi-automated process that necessitates human intervention and ML expertise to construct an effective pipeline and select an appropriate GNN method. Data scientists may choose the most appropriate GNN method based on various constraints, such as time or memory limitations. Furthermore, as depicted in Figure~\ref{fig:GNN_Pipeline}, the separation of the trained models from the KG engines adds an extra layer of complexity for data scientists to apply their models when inferring the KG.\shorten

\section{Challenges of GML-enabled KG Engine}
\label{sec:ChallengesOfAutomatingCognitiveQuery}

This section highlights the open research challenges and opportunities raised by developing GML-enabled KG Engine.


\subsection{Automatic Training: Method Selection and Meta-sampling}
\label{subSec:KG_GML_support}

\begin{figure}[t]
\ncp\ncp
  \centering  \includegraphics[width=\columnwidth,draft=false]{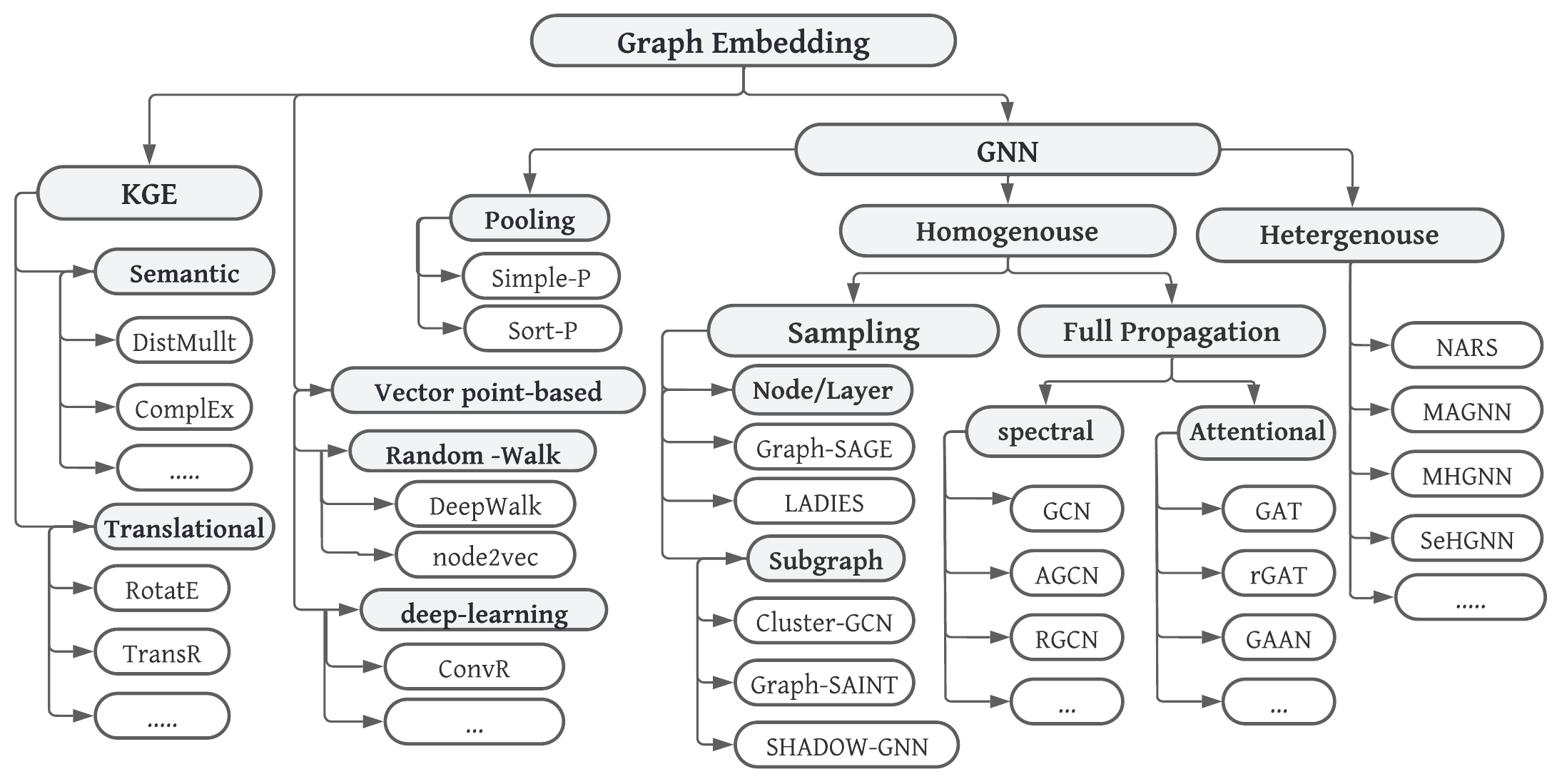}   
  \ncp\ncp\ncp\ncp\ncp\ncp
  \caption{A taxonomy of methods for training GML models.}   
  \label{fig:EmbeddingsChart}
  \ncp\ncp\ncp
\end{figure}

There are numerous methods for training models for GML tasks, as summarized in Figure~\ref{fig:EmbeddingsChart}. These methods could be classified mainly into two categories KG embeddings (KGE) or graph neural network (GNN) methods. Examples of KGE methods are TransE, RotatE, ComplEx, and DistMult
~\cite{KG-Embedding-Survey}. Some GNN methods support sampling on full graph, such as Graph-SAINT\cite{GraphSAINT}, Shadow-SAINT \cite{Shadow-GNN}, and MorsE\cite{MorsE}. Examples of GNN  full-batch training (without sampling) methods are RGCN \cite{RGCN} and GAT\cite{r-GAT}. Our taxonomy has more categories, as shown in Figure~\ref{fig:EmbeddingsChart}.\shorten

GML methods vary significantly in terms of their accuracy, training time, and memory requirements. Furthermore, the complexity of each GML task may differ depending on various factors, such as the size of KGs and the number of node/edge types related to the task. For example, link prediction can be more resource-intensive than node classification.
Different GML methods may perform differently under the same budget constraints, and selecting the best method can depend on several factors. Hence, automating a training pipeline for a specific GML task based on a user's budget for time and memory is challenging. For instance, some GML methods perform full-batch training, which requires more memory budget. 
These methods require huge memory to train models on large KGs. Some other GNN methods may suffer from over-smoothing, which can cause accuracy degradation. Sampling-based GNN (mini-batch training) methods use different types of sampling, which vary in avoiding these limitations. Therefore, automating the selection of GML methods for a specific task based on a given time or memory budget is challenging.  

Real KGs can contain millions to billions of triples, such as DBLP\cite{KG_DBLP} and MAG~\cite{MAG}. However, training GML models on these large KGs requires colossal computing resources that exceed the capabilities of a single machine. As a result, there is a need for identifying a smaller training dataset of the KG, which is specific to the task at hand. This process is known as meta-sampling. It has been proposed in various application domains, including computer vision~\cite{PointCloudMetaSampling, Learning_to_Sample} and speech recognition~\cite{ASR_Meta_Sampling}, to extract a training dataset that is tailored to the given task. In the context of GML, meta-sampling presents an opportunity to optimize training models on large KGs by selecting a representative sub-graph that is relevant to the task. This approach can help reduce time and memory requirements without sacrificing accuracy. Therefore, exploring the potential benefits of using meta-sampling in training GML models to extract task-specific subgraphs is crucial. By doing so, we can improve the efficiency and effectiveness of GML methods on large-scale KGs. This raises a research opportunity to explore different meta-sampling approaches for GML methods on large knowledge graphs (KGs).\shorten  

\begin{figure*}[t]
\ncp\ncp\ncp
  \centering
  \includegraphics[width=\textwidth,draft=false]{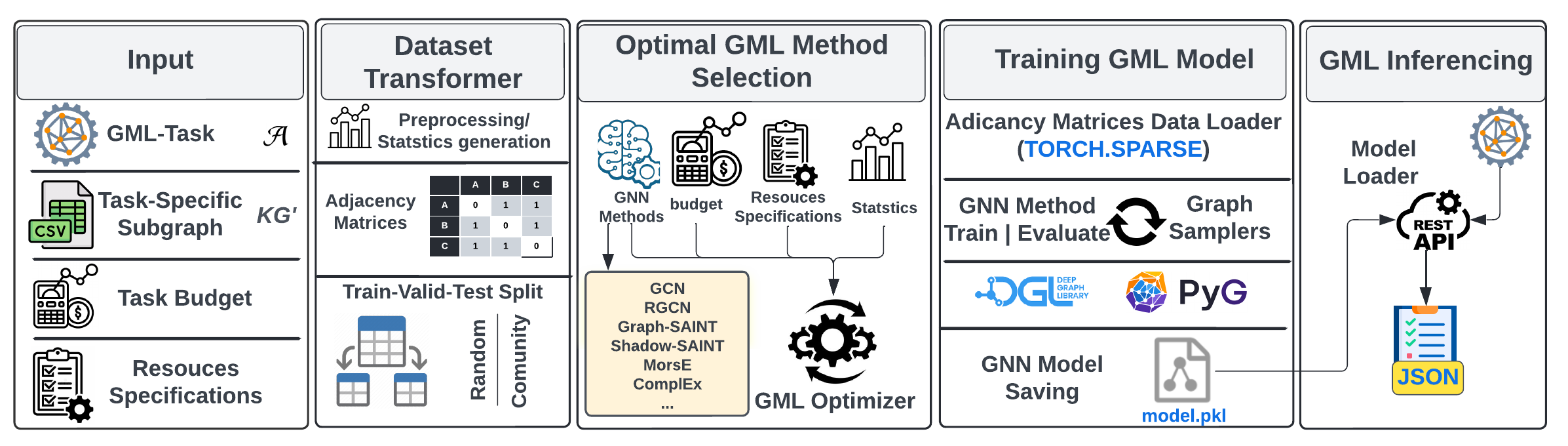}
  \ncp\ncp\ncp\ncp\ncp
  \caption{The automation of training pipeline and inference in our GML-as-a-service (GMLaaS). GMLaaS interacts with the KGMeta Manager to train a model for a specific task with limited budget. The automated pipeline opt to the near-optimal GML method for training a model within a limited budget. GMLaaS supports task inference through RestAPI that is called by a UDF.\shorten}   
  \label{fig:kgnet_training_pipeline}  
  \ncp\ncp\ncp\ncp
\end{figure*}

\subsection{Seamless Integration Between GML Models and KGs}
\label{subSec:Support_RDF_engine_with_semantic_operators}
Enabling GML on top of RDF engines poses significant challenges, mainly interfacing between the trained models and the underlying data management engine. One common approach is to use user-defined functions (UDFs) to implement this interface~\cite{ML_Infernce, bordawekar2016enabling, In-Database-ML}. However, this comes with a cost for query optimizations in data systems~\cite{CostEA_UDFS}.
The existence of an extensive catalog of UDFs can limit the expressiveness of ML-based queries. For instance, a large catalog of UDFs makes it difficult for users to choose between UDFs and find the right one for their needs. Most existing query optimizers do not have models estimating the cost of these UDFs. Hence, automating the query optimization of SPARQL$^\mathbf{ML}$ queries is challenging.  
There is a research opportunity for seamless integration between trained GML models and  RDF. To address these challenges, we proposed KGMeta as a graph representation of metadata of trained models interlinked with the KGs.\shorten

\subsection{Optimizing SPARQL$^\mathbf{ML}$ Queries and Benchmarks } 

User-defined predicates were first proposed for SQL~\cite{ChaudhuriS99}. In SPARQL$^\mathbf{ML}$, a user-defined predicate is used to get a prediction from one of the trained models associated with a specific node in the graph. Estimating the cost of evaluating a user-defined predicate is more complex than estimating the cost of a traditional RDF predicate. While cardinality estimation is used to optimize only the execution time for the latter, a user-defined predicate in a SPARQL$^\mathbf{ML}$ query can be inferred by multiple models, each with varying accuracy and inference time. RDF engines are unaware of this information, leading to the problem of selecting the best model for inference.\shorten   

For a SPARQL$^\mathbf{ML}$ query, the inference step in an RDF engine using a chosen model is a challenging task that requires optimization, specifically for rank-ordering the inference process. The challenge lies in deciding whether to perform the inference in a single call to a UDF or per instance, which may result in an extensive number of UDF calls. Additionally, each model has a unique cardinality, i.e., the total number of predictions it can make. This makes predicting rank-ordering complex as RDF engines lack accurate estimation of UDF costs.\shorten

To address these challenges, there are research opportunities for developing benchmarks to evaluate optimization approaches for SPARQL$^\mathbf{ML}$ queries. These benchmarks should consider various models for different user-defined predicates and be designed to work with large datasets. Furthermore, each SPARQL$^\mathbf{ML}$ query should vary in the number of user-defined predicates and be associated with variables of different cardinalities. This will enable a comprehensive evaluation of the performance and scalability of varying optimization approaches for SPARQL$^\mathbf{ML}$ queries.\shorten

\section{The {\sysName} Platform}
\label{sec:arch}

{\sysName} provides two main services, namely GML as a Service (GMLaaS) and SPARQL$^\mathbf{ML}$ as a Service on top of existing RDF engines, as shown in Figure~\ref{fig:sys}. 

\subsection{GML as a Service (GMLaaS)}
\label{kgnet_gml_as_service}

{\sysName} is a platform that offers end-to-end automation of GML training on KGs, as depicted in Figure~\ref{fig:kgnet_training_pipeline}. The platform provides \emph{GMLaaS}, a Restful service that manages GML models in terms of automatic training and interactive inferencing. Additionally, it utilizes an embedding store to facilitate entity similarity search tasks by computing the similarity between embedding vectors. The \emph{GML training manager} automates the training pipeline per task. However, the automation of GML training on KGs is challenging due to the complexity and size of KGs. Therefore, {\sysName} leverages our meta-sampling approach to optimize the training process by selecting a task-specific subgraph ($KG'$) that is specific to the given task. This step helps reduce the time and memory required without trading accuracy. The pipeline takes as input a task-specific subgraph ($KG'$), the GML task, the task budget, and the available resources within the ML environment. 

The \emph{Data Transformer} step converts the subgraph into a sparse-matrix format optimized for in-memory and matrix operations. This format is compatible with popular graph ML data loaders, such as Py-Geometric and DGL, and is ideal for sparse KGs. Our pipeline ensures data consistency by validating node/edge types counts, removing literal data and target class edges, and generating graph statistics. We also perform a train-validation-test split using different strategies like random and community-based. {\sysName} automates this transformation, making ad-hoc GML training queries possible.

\begin{figure}[t]
  \centering
  \includegraphics[width=\columnwidth,draft=false]{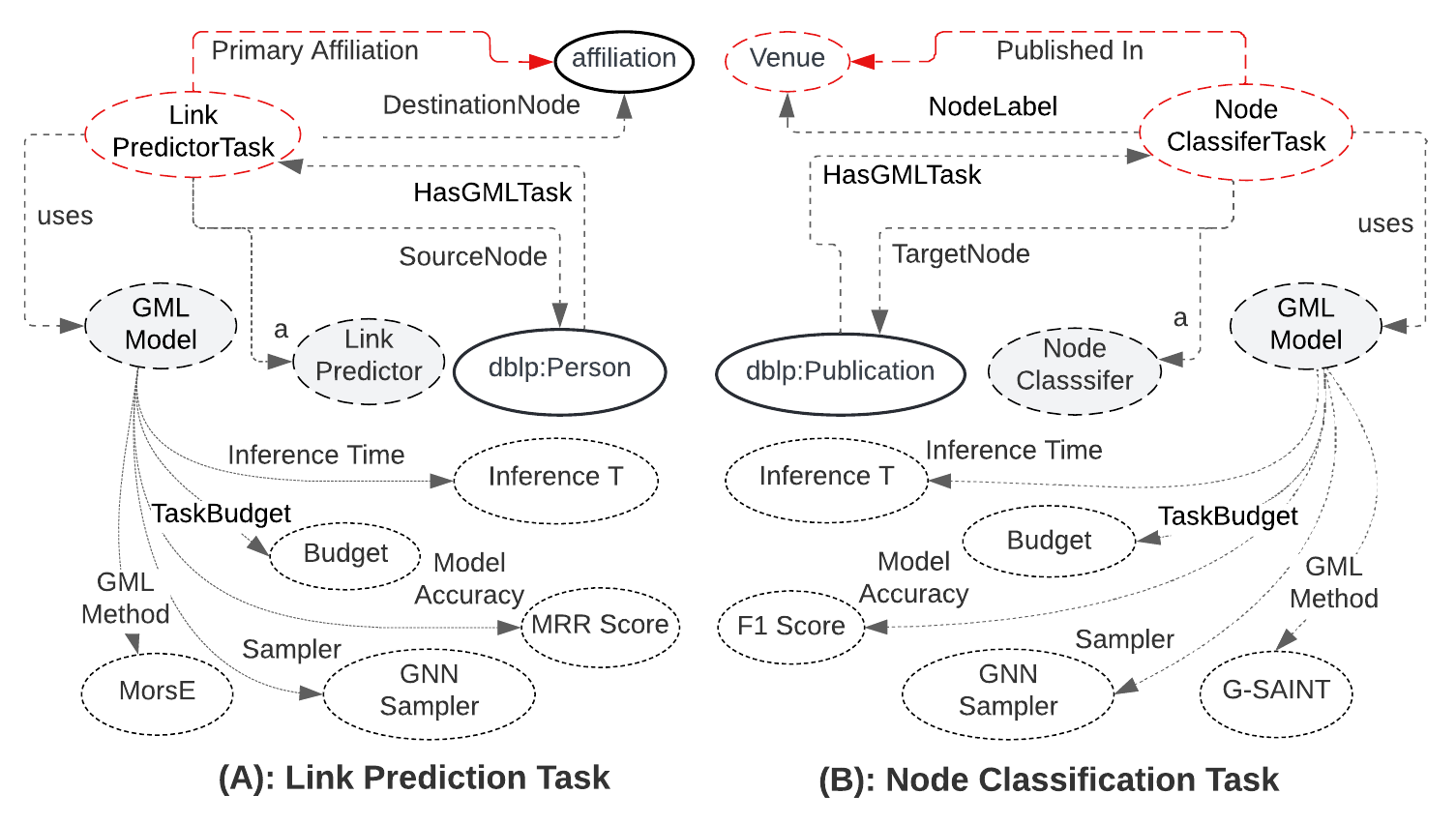}
  \caption{A KGMeta graph of two trained models for node classification and link prediction tasks. The white nodes are nodes from the original data KG. The dashed nodes/edges are metadata collected per trained model.}  
  \label{fig:Meta-data_KG_Schema}
  \ncp\ncp\ncp\ncp
\end{figure}

\begin{figure}[t]
\lstdefinestyle{myStyle}{
  basicstyle=\small\ttfamily,
  language=SPARQL,
   }
\begin{lstlisting}[style=myStyle,captionpos=b,showspaces=false,frame=singlem,numbers=left,xleftmargin=0.5cm,morekeywords={kgnet,dblp, NodeClassifer,Insert},escapechar=\%]
prefix dblp:<https://www.dblp.org/>
prefix kgnet:<https://www.kgnet.com/>
Insert into <kgnet>  { ?s ?p ?o }
where {select * from  %\textbf{\textcolor{purple}{kgnet.TrainGML}(}%
{Name: 'MAG_Paper-Venue_Classifer',
 GML-Task:{ TaskType:%\textbf{kgnet:NodeClassifier}%,
 TargetNode: dblp:publication, 
 NodeLable: dblp:venue},
 Task Budget:{ MaxMemory:50GB, MaxTime:1h,
               Priority:ModelScore} } %\textbf{)}%}; 
\end{lstlisting}
\caption{A SPARQL$^\mathbf{ML}$ insert query that trains a paper-venue classifier on DBLP. The TrainGML function is a UDF that is implemented inside the RDF engine.}
\label{fig:paperVenueTrainGML}
\ncp\ncp\ncp\ncp\ncp\ncp\ncp\ncp\ncp
\end{figure}


The \emph{Optimal GML Method Selection} step selects the best GML method for a given task. {\sysName} supports various GNN methods, including GCN, RGCN, Graph-SAINT, Shadow-SAINT, Morse, and KGE methods such as ComplEx. We estimate the required memory for each method based on the size and the number of generated sparse-matrices, as well as the training time based on the matrix dimensions and feature aggregation approach. Moreover, we estimate the training time based on the dimension of the sparse-matrices and GNN neighbour nodes features aggregation approach adopted by each method. For GNN sampling-based methods, the sampling cost basically depends on the sampling heuristic used \cite{Serafini21}. Thus, we are working on a more advanced estimation method based on sampling the sparse-matrices and running a few epochs on them.

{\sysName}'s GML-optimizer determines the necessary resources for each method and optimizes the training settings, ensuring scalability in distributed environments. The automated pipeline trains a model and collects evaluation metrics and inference time statistics. A URI is generated for the trained model to distinguish it from other models used for inference tasks. The model meta-data is returned to the KGMeta Manager to update the KGMeta graph. Figure \ref{fig:Meta-data_KG_Schema}.a and b show the generated meta-data for link prediction and node classification models, respectively. The \emph{Embedding Store} sub-component, shown in Figure~\ref{fig:sys}, is used for fast similarity search by storing, indexing, and searching embeddings. The \emph{GML Inferencing} receives HTTP calls for inference, serializes the result into a JSON Restful-API response, and sends it back to the RDF engine, as shown in Figure~\ref{fig:sys}. The current version uses FAISS embedding store~\cite{faiss2017} to enable ad-hoc queries for node similarity search.


 \begin{figure}[t]
\lstdefinestyle{myStyle}{
  basicstyle=\small\ttfamily,
  language=SPARQL,
   }
\begin{lstlisting}[style=myStyle,captionpos=b,showspaces=false,frame=singlem,numbers=left,xleftmargin=0.5cm,morekeywords={kgnet,dblp, NodeClassifer,delete},escapechar=\%]
prefix dblp:<https://www.dblp.org/>
prefix kgnet:<https://www.kgnet.com/>
%\textbf{\textcolor{purple}{delete \{?NodeClassifier ?p ?o\}}}%
where {
?NodeClassifier a kgnet:NodeClassifier.
?NodeClassifier kgnet:TargetNode dblp:Publication.
?NodeClassifier kgnet:NodeLabel dblp:venue.}
\end{lstlisting}
\caption{A SPARQL$^\mathbf{ML}$ delete query that deletes a trained model and its meta-data.}
\label{fig:GMLDeleteQuery}
\end{figure}

\begin{figure}[t]
\lstdefinestyle{myStyle}{
basicstyle=\small\ttfamily,
  language=SPARQL,
}
\begin{lstlisting}[style=myStyle,captionpos=b,showspaces=false,frame=singlem,numbers=left,xleftmargin=0.5cm,morekeywords={kgnet,dblp, LinkPredictor,select},escapechar=\%]
prefix dblp: <https://www.dblp.com/>
prefix kgnet: <https://www.kgnet.com/>
select ?author ?affiliation
where { ?author a dblp:person.
?author ?LinkPredictor ?affiliation.
?LinkPredictor a %\textcolor{purple}{ \textbf{kgnet:LinkPredictor}}%.
?LinkPredictor kgnet:SourceNode dblp:person.
?LinkPredictor kgnet:DestinationNode dblp:affiliation.
?LinkPredictor kgnet:TopK-Links 10.}
\end{lstlisting}
\ncp\ncp\ncp\ncp
\caption{A SPARQL$^\mathbf{ML}$ query predicting author affiliation link (edge) on DBLP KG.}
\label{fig:AuthorAffaliationSPARQL}
\ncp\ncp\ncp
\end{figure}


\subsection{The SPARQL$^\mathbf{ML}$ as a Service}
\label{kgnet_Query_Manager}

We offer a SPARQL$^\mathbf{ML}$ as a Service, which comprises three main components: Query Manager, KGMeta Governor, and Meta-sampler. In addition, we provide an interfacing language called SPARQL$^\mathbf{ML}$ that enables users to express SPARQL-like queries for INSERT, DELETE, or SELECT operations, such that: \myNum{i} a SPARQL$^\mathbf{ML}$ \textit{INSERT} query is used to train a GML model and maintain its metadata in KGMeta (as shown in Figure~\ref{fig:paperVenueTrainGML}), \myNum{ii} a SPARQL$^\mathbf{ML}$ \textit{DELETE} query is used to delete trained model files and associated embeddings from the GML-aaS component and then deletes its metadata from the KGMeta (as in Figure~\ref{fig:GMLDeleteQuery}), \myNum{iii} a SPARQL$^\mathbf{ML}$ \textit{SELECT} query is for querying and inferencing the KG, e.g., the query in Figure~\ref{fig:AuthorAffaliationSPARQL}. When a SPARQL$^\mathbf{ML}$ query is received, the Query Manager parses it. An INSERT or DELETE query is sent to the KGMeta Governor. If it is a SELECT query, it is optimized and rewritten as a SPARQL query.\shorten 

\subsubsection{KGMeta Governor}
The KGMeta Governor maintains a KGMeta graph for each KG, using statistics and metadata collected from trained GML models specific to that KG. The INSERT query is a request to train a task on a certain KG. The parsed information includes the task type (such as node classification or link prediction), the task inputs (such as the target nodes and classification labels (Y classes) for a classification task), and a budget (such as memory and time budget). Experienced ML users can provide additional information, such as hyperparameters or a specific GML method. This information is encapsulated as a JSON object, as shown in Figure~\ref{fig:paperVenueTrainGML}. At line 4, the \emph{TrainGML} is a UDF that takes as input a JSON object that encapsulates all required information to train a GML model. The KGMeta Governor sends the task to the meta-sampler to obtain a task-specific subgraph ($KG’$) for the given task. Then governor interacts with the GML Training Manager to automate the training pipeline for this task. Once training is complete, the KGMeta Governor receives the trained model's metadata, including accuracy and inference time, to maintain the KGMeta, as illustrated in Figure~\ref{fig:Meta-data_KG_Schema}.

\begin{figure}[t]
\lstdefinestyle{myStyle}{basicstyle=\small\ttfamily, language=SPARQL }
\begin{lstlisting}[style=myStyle,captionpos=b,showspaces=false,frame=singlem,numbers=left,xleftmargin=0.5cm,morekeywords={kgnet, getNodeClass,getKeyValue,dblp},escapechar=\%]
prefix dblp: <https://www.dblp.org/>
prefix kgnet: <https://www.kgnet.com/>
select ?title 
       sql:UDFS.getNodeClass($m,?paper) as ?venue 
where { 
    ?paper a dblp:Publication.
    ?paper dblp:title ?title.
}
\end{lstlisting}
\ncp\ncp\ncp\ncp
\caption{A candidate SPARQL for SPARQL$^\mathbf{ML}_{pv}$}
\label{fig:SPARQL1}
\ncp\ncp\ncp\ncp\ncp\ncp
\end{figure}

\subsubsection{Meta-sampler}
Our meta-sampler aims to identify a task-specific subgraph ($KG'$) for training a GML task. Each GML task targets nodes of a specific type, such as dblp:Publication in SPARQL$^\mathbf{ML}_{pv}$. Our meta-sampler extracts a task-specific subgraph ($KG'$), which comprises a set of triples with representative triples associated with the target nodes. Based on the KG schema structure the size of $KG'$ is much smaller than the size of KG. This smaller size will optimize training time and require less memory for training the GML task $\mathcal{A}$. However, the KG may contain triples that are not reachable from a target node $v^T$ or connected via more than three hops from $v^T$. These triples do not assist the model in generalizing and may lead to over-smoothing problems~\cite{GNN_Over_Smoothing,EIGNN_Oversmoothing}. \shorten

Our SPARQL-based meta-sampling method determines the scope of the extracted subgraph based on two parameters: \myNum{i} the direction $d$, where $d = 1$ for outgoing and $d = 2$ for bidirectional (i.e., both outgoing and incoming), and \myNum{ii} the number of hops $h$. We evaluated the performance of our method using four combinations of $d \in \{1, 2\}$ and $h \in \{1, 2\}$. Our meta-sampling approach achieved better results with $d=1$ and $h=1$ for node classification, whereas for link prediction, our meta-sampling method performed better with $d=2$ and $h=1$.\shorten

\subsubsection{The Query Manager} 
The Query Manager is responsible for optimizing SPARQL$^\mathbf{ML}$ queries for model selection and rank-ordering to evaluate user-defined predicates. In the case of SPARQL$^\mathbf{ML}_{pv}$ shown in Figure \ref{fig:paperVenueSPARQL}, the query optimizer fetches all URIs of the models satisfying the conditions associated with the user-defined predicate \emph{?NodeClassifier}. The KGMeta is an RDF graph containing optimizer statistics, such as model accuracy, inference time, and model cardinality. Therefore, we use a SPARQL query to obtain the models' URIs, accuracy, inference time, and cardinality. The query optimizer selects the near-optimal GML model that achieves high accuracy and low inference time. We define this problem as an integer programming optimization problem to minimize total execution time or maximize inference accuracy.

The \emph{SPARQL$^\mathbf{ML}$ Query Re-writer} uses the near-optimal GML model with URI $m$ to generate a candidate SPARQL query. {\sysName} currently supports two possible execution plans, whose query templates are shown in Figures~\ref{fig:SPARQL1} and~\ref{fig:SPARQL2}. The core idea is to map a user-defined predicate into a user-defined function (UDF), such as \emph{sql:UDFS.getNodeClass}, to send HTTP calls during the execution time to the GML Inference Manager in our GMLaaS to get inference based on the pre-trained model $m$. The number of HTTP calls may dominates the query execution cost. For example, SPARQL$^\mathbf{ML}_{pv}$ predicts the venue of all papers, whose size is |?papers|.

\begin{figure}[t]
\lstdefinestyle{myStyle}{basicstyle=\small\ttfamily, language=SPARQL }
\begin{lstlisting}[style=myStyle,captionpos=b,showspaces=false,frame=singlem,numbers=left,xleftmargin=0.5cm,morekeywords={kgnet, getNodeClass,getKeyValue,dblp},escapechar=\%]
prefix dblp: <https://www.dblp.org/>
prefix kgnet: <https://www.kgnet.com/>
select ?title 
  sql:UDFS.getKeyValue(?venues_dic,?paper) as ?venue 
where { 
    ?paper a dblp:Publication.
    ?paper dblp:title ?title.
{select sql:UDFS.getNodeClass($m,dblp:Publication)
  as ?venues_dic where { } }}
\end{lstlisting}
\ncp\ncp\ncp\ncp
\caption{A candidate SPARQL for SPARQL$^\mathbf{ML}_{pv}$}
\label{fig:SPARQL2}
\ncp\ncp\ncp
\end{figure}


The query template shown in Figure~\ref{fig:SPARQL1}  will generate |?papers| HTTP calls. However, the query template shown in Figure~\ref{fig:SPARQL2} reduces the number of HTTP calls to one by enforcing an inner select query constructing a dictionary of all papers and their predicted venues. Then, \emph{sql:UDFS.getKeyValue} is used to look up the venue of each paper. Our query optimizer decomposes the triple patterns related quering the KG triples in the SPARQL$^\mathbf{ML}$ query into sets per variable associated with a user-defined predicate. For example, in the SPARQL$^\mathbf{ML}_{pv}$ query shown in Figure~\ref{fig:paperVenueSPARQL}, our optimizer identifies two triple patterns that match the variable \emph{?paper} and one triple pattern that matches the variable \emph{?venue}. We use a SPARQL query to get the cardinality of each set, which is the number of distinct values of the variable in the dataset. We formulate this problem as another integer programming optimization problem~\cite{AppliedMathematicaProgramming} that minimizes the total number of HTTP calls or minimizes the constructed dictionary size, which is based on the model cardinality. For instance, in the query shown in Figure~\ref{fig:SPARQL2}, our optimizer generates a dictionary of all papers and their predicted venues, which is then used to retrieve the venue of each paper using the UDF \emph{sql:UDFS.getKeyValue}.

\section{Experimental Evaluation}   
\label{sec:eval}

This section analyzes the ability of {\sysName} in automating pipelines to train a model for a specific task with less time and memory w.r.t traditional pipelines on full graphs.

\subsection{Evaluation Setup}

\textbf{Compared Methods:} 
We used  RGCN \cite{RGCN} as a full-batch training method and GraphSAINT \cite{GraphSAINT}, ShadowSAINT \cite{Shadow-GNN} as mini-batch sampling-based methods for node classification and MorsE\cite{MorsE} as edge sampling-based method for link prediction. The OGB \cite{OGB} default conﬁgurations are used in both sampling and training. 
Node features are initialized randomly using Xavier weight initialization in all experiments. 
 \begin{table}[t]
 \centering
   \caption{Statistics of the used KGs and GNN tasks. We used four times larger KGs (DBLP and Yago) than the ones reported in OGB~\cite{OGB}.\shorten}
  \label{tbl_GNN_usecases}
  \begin{tabular}{lrr}
    \hline
    \textbf{Knowledge Graph}&DBLP&YAGO4\\
    \hline
    \textbf{\#Triples} &252M& \hfil400M \\
    \textbf{\#Targets}& \makecell[r]{   50 Venue \\ 51K Affiliations \\ 1.2M paper}&200 Country\\
    \textbf{\#Edge Types}&48 &98 \\
    \textbf{\#Node Types}&42 &104 \\
    \textbf{Tasks}&NC,LP,ES &NC \\
    \hline            
\end{tabular}
\end{table}


\textbf{Computing Infrastructure:}
All experiments are conducted on Ubuntu server virtual machine that is equipped with dual 64-core Intel Xeon 2.4 GHZ (Skylake, IBRS) CPUs, 256 GB of main memory and 1TB of disk storage. 

\textbf{Real KGs:}
We mainly focus on two benchmark KGs distinguishing in graph size, graph data domain, task type, and connection density including (DBLP\cite{KG_DBLP}  and Yago-4 \cite{KG_Yago4}). We conducted two node classification tasks and one link prediction. We followed the tasks used in OGB~\cite{OGB}. Statistics about used KG and tasks are provided in Table~\ref{tbl_GNN_usecases}.\shorten
    
\textbf{Endpoints:}
We use Virtuoso  07.20.3229 as a SPARQL endpoint, as it is widely adopted as an endpoint for large KGs, such as DBLP. The standard, unmodified installation of the Virtuoso engine was run at the endpoints and used in all experiments.\shorten

  \begin{figure}[t]
  \centering  \includegraphics[width=\columnwidth,draft=false]{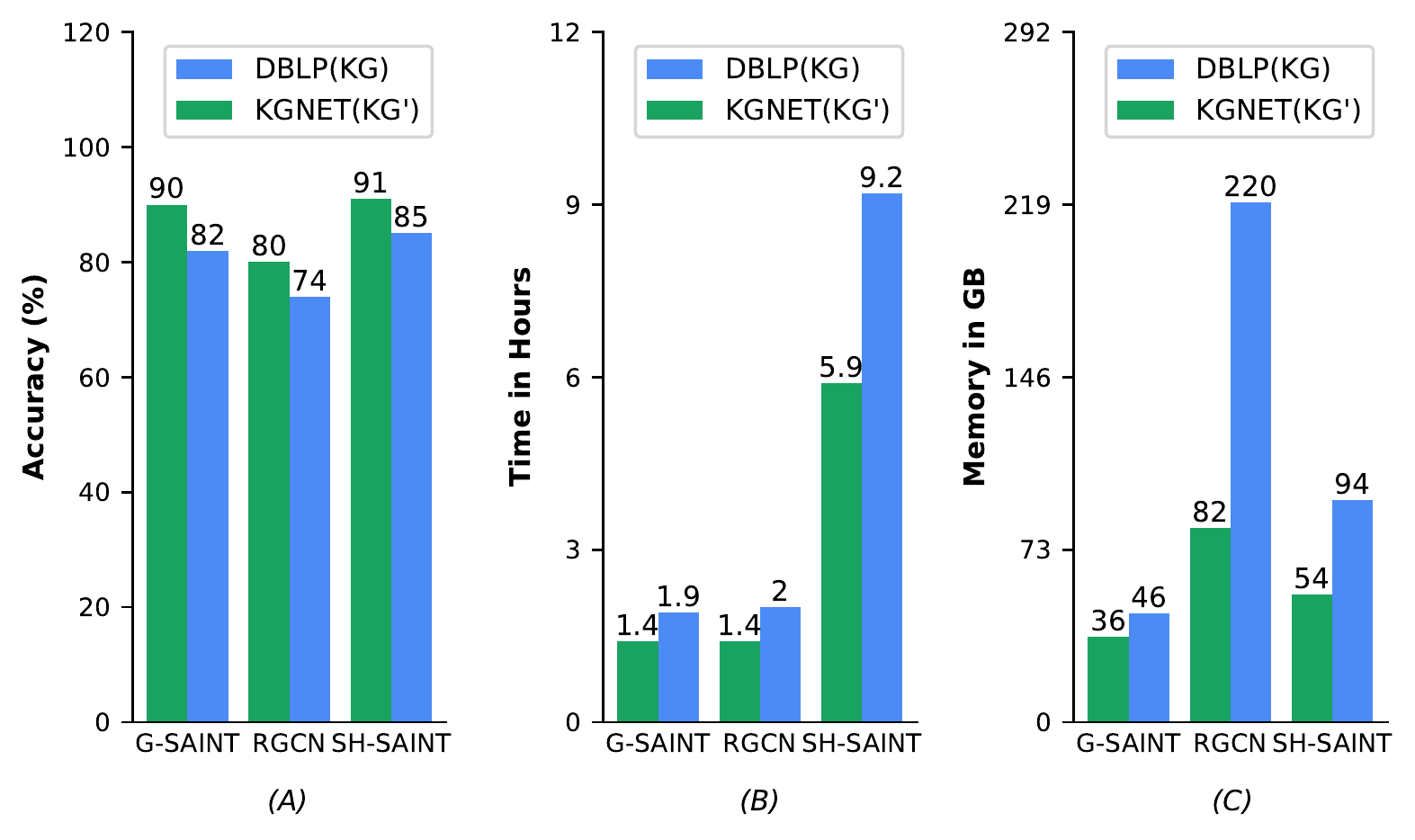} 
 \vspace*{-2ex}
  \caption{
    (a) Accuracy, (B) Training Time, and (C) Training Memory for DBLP KG Paper-Venue node classification task. The {\sysName} task-oriented sampled subgraph (KG')  significantly improves accuracy, training time, and memory.
  }   
  \ncp\ncp\ncp\ncp\ncp\ncp
  \label{fig:KG_sampling_NC_DBLP} 
\end{figure} 


\begin{figure}[t]
  \centering  \includegraphics[width=\columnwidth,draft=false]{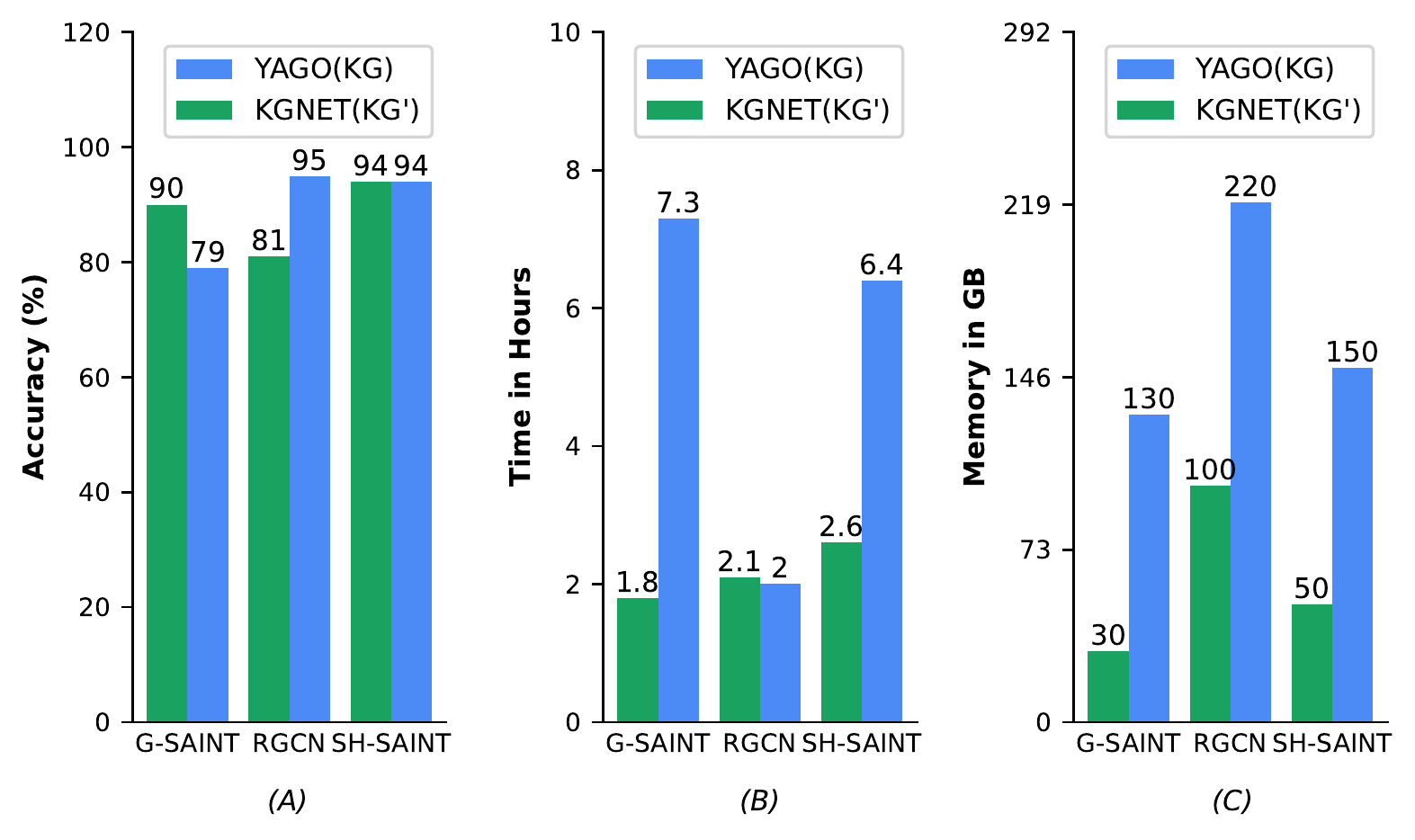} 
  \vspace*{-3ex}
  \caption{
    (a) Accuracy, (B) Training Time, and (C) Training Memory for YAGO-4 KG Place-Country node classification task. The {\sysName} task-oriented sampled subgraph (KG') significantly improves accuracy, training time, and memory.
  }   
  
  \label{fig:KG_sampling_NC_YAGO} 
\end{figure} 
 \begin{figure}[t]
  \centering  \includegraphics[width=\columnwidth,draft=false]{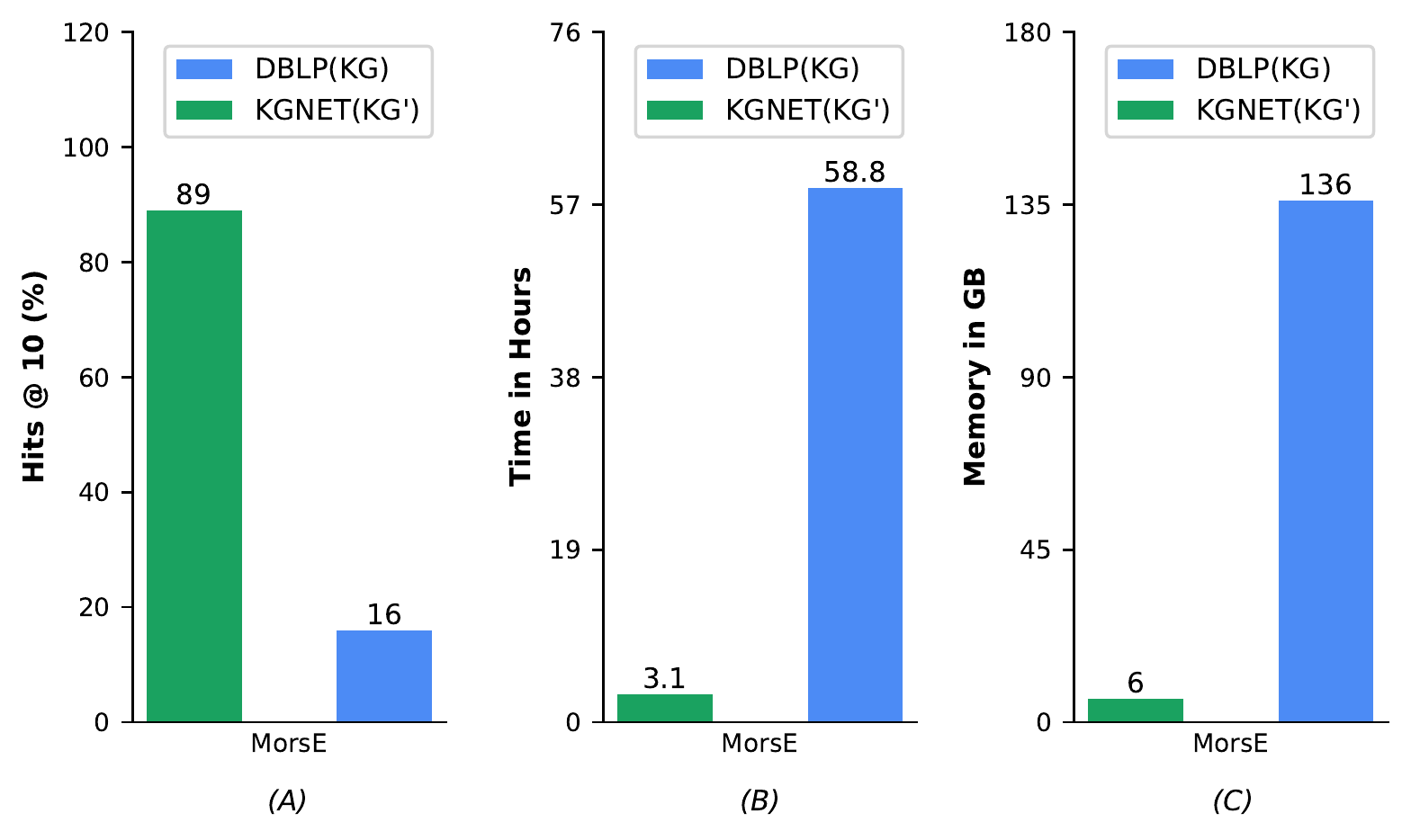} 
  \vspace*{-3ex}
  \caption{
    (a) Accuracy, (B) Training Time, and (C) Training Memory for the DBLP Author Affiliation link prediction task. The {\sysName} task-oriented edge sampled subgraph (KG') significantly improves the Hits@10 MRR score, training time, and training memory.\shorten
  }   
  \ncp\ncp\ncp
  \label{fig:KG_sampling_LP_DBLP-AA} 
\end{figure}

\subsection{GML Experiments With Real KGs}

Three GML tasks are conducted to evaluate the {\sysName} automated GML pipeline. For \textbf{Node classification task}, GNN methods are used to train node classifiers to predict a venue for each DBLP paper. The KG is loaded into the Virtuoso RDF engine. 
{\sysName} performs meta-sampling using $d1h1$ to extract the task-specific subgraph ($KG'$) to train RGCN, Graph-SAINT, and Shadow-SAINT methods. The task results in Figures~\ref{fig:KG_sampling_NC_DBLP} and~\ref{fig:KG_sampling_NC_YAGO} show that our KGNet training pipeline using ($KG'$) outperforms the traditional pipeline on the full KG in all methods with up to 11\% accuracy score. The automated training pipeline of {\sysName} has successfully enabled GNN methods to achieve significant reductions in memory consumption and training time. Specifically, {\sysName} has demonstrated a reduction of at least 22\% in memory consumption and 27\% in training time. These results demonstrate that {\sysName} can effectively discover task-specific subgraphs for each task.\shorten

Our \textbf{Link prediction task} aims to predict an author's affiliation link based on their publications and affiliations history on the DBLP knowledge graph. MorsE~\cite{MorsE} is the state-of-the-art link-prediction sampling-based method. We use the MorsE in the traditional pipeline with the full KG. In the {\sysName} pipeline,  our meta-sampling first extracts the task-specific subgraph ($KG'$) using $d2h1$ to train MorsE. 
The results, shown in Figure~\ref{fig:KG_sampling_LP_DBLP-AA}, demonstrate that the KGNet automated pipeline outperforms the pipeline trained on the full KG in terms of Hits@10 MRR score. {\sysName} achieves a significant reduction in memory usage and training time, with a reduction of 94\% compared to the pipeline trained on the full KG.\shorten

\section{Related Work} 
\label{sec:related_work}

The adoption of combining AI and database systems has been growing rapidly, with two main approaches emerging: AI models incorporated in DB systems (AI4DB) and database techniques optimized for AI systems for better scalability (DB4AI)~\cite{SQL-ML-Survey}. In {\sysName}, we classify SPARQL$^\mathbf{ML}$ as part of the AI4DB approach since we have extended the KG engine to query and perform inference on KGs using GML models. However, we classify GMLaaS as part of the DB4AI approach since we have optimized the training pipeline using our meta-sampling approach, which queries a KG to extract a task-specific subgraph. 
Works RDFFrames \cite{RDFFrames}, DistRDF2ML\cite{DistRDF2ML}, and Apple Saga \cite{Saga_Apple}  aim to bridge the gap between ML and RDF systems by enabling the user to extract data from heterogeneous graph engines in a standard tabular format to apply traditional ML tasks such as classification, regression, and clustering or use KGE methods to generate node/edge embeddings for similarity search applications.

Yuo Lu et.al. addressed the problem of AI-enabled query optimization for SQL in~\cite{ML_Infernce} and introduced the probabilistic predicates (PPs) method that can be trained without any knowledge of the inference models. In Learned B+tree~\cite{B+tree_ML}, the B+tree index is optimized based on AI models that map each query key to its page. Hasan et al~\cite{Multi-Attribute-Selectivity} allow fast join queries by utilizing auto-regressive densities model to represent the joint data distribution among columns. ITLCS~\cite{IndexSelection_ML} introduced an index selection ML-based method that uses a classiﬁer model as well as a genetic algorithm that selects the accurate index. Stardog~\cite{Stardog-ML} supports supervised learning to build predictive analytics models. Stardog enables users to write SPARQL queries that collect the ML training features set in a tabular format and apply classical ML, i.e., classification, clustering, and regression that can be used for inference queries.\shorten

Google's BigQuery ML~\cite{BigQuery-ML} provides user-friendly tools to support AI models in SQL statements by introducing a hybrid language model that contains both AI and DB operations, which executes AI operations on AI platforms such as TensorFlow and Keras. SQL4ML\cite{sql4ml} translates ML operators implemented in SQL into a TensorFlow pipeline for efficient training. To enable ad-hoc GML pipelines using SPARQL, RDF engines require this level of support.

Bordawekar et al. \cite{bordawekar2017cognitive} built a cognitive relation database engine that queries database records utilizing word similarity using word2vec embeddings and extends results with external data sources. The cognitive DB represents a step towards linking representation learning with DB using text embedding techniques. EmbDI~\cite{EmbDI} automatically learns local relation embeddings with high quality from relational datasets using a word embedding to support datasets schema matching. Unlike all the above-mentioned systems, {\sysName} proposed a platform combining DB4AI and AI4DB approaches to bridge the gap between GML frameworks and RDF engines.



\section{Conclusion}
\label{sec:conc}

The lack of integration between GML frameworks and RDF engines necessitates that data scientists manually optimize GML pipelines to retrieve KGs stored in RDF engines and select appropriate GML methods that align with their computing resources. Furthermore, the trained models cannot be directly used for querying and inference over KGs, which impedes systems' scalability, particularly as KGs grow in size and require excessive computing resources. Additionally, these limitations impact the system's flexibility, as descriptive query languages are incapable of incorporating GML models. 
To overcome these limitations, this vision paper proposed {\sysName}, 
an on-demand GML-as-a-service (GMLaaS) platform on top of RDF engines to support GML-enabled SPARQL queries (SPARQL$^\mathbf{ML}$). {\sysName} uses meta-sampling to extract a task-specific subgraph ($KG'$) as a search query against a KG for a specific task. Our GMLaaS automates a cost-effective pipeline using $KG'$ to train a model within a given time or memory budget. 
{\sysName} maintains the metadata and statistics of trained models as an RDF graph called KGMeta, which is stored alongside associated KGs. KGMeta leads to a seamless integration between GML models and RDF engines, allowing users to easily express their SPARQL$^\mathbf{ML}$ queries based on the SPARQL logic of pattern matching. Moreover, KGMeta enables {\sysName} to optimize SPARQL$^\mathbf{ML}$ queries for model selection and rank-ordering for the inferencing process. {\sysName} raises research opportunities spanning across data management and AI.\shorten

\balance
\bibliographystyle{IEEEtran}
\bibliography{references}

\end{document}